\documentclass[twocolumn,pre,aps,showpacs]{revtex4}
\usepackage{graphicx}
\usepackage{amsmath}
\def\bear{\begin{eqnarray}}
\def\eear{\end{eqnarray}}
\newcommand{\Fig}[1]{Fig.~\ref{#1}}
\newcommand{\Figs}[1]{Figs.~\ref{#1}}
\newcommand{\Eqn}[1]{Eq.~(\ref{#1})}
\newcommand{\Eqns}[1]{Eqs.~(\ref{#1})}
\newcommand{\Sec}[1]{Sec. \ref{#1}}

\begin{document}
\bibliographystyle{revtex}

\title{On the conditions for synchronization in unidirectionally
coupled chaotic oscillators}

\author{
P.~Muruganandam%
}
\email{anand@bdu.ernet.in}
\affiliation{Centre for Nonlinear Dynamics, Department of Physics
Bharathidasan University, Tiruchirapalli 620 024, India}
\author{
S.~Parthasarathy%
}
\email{spartha@rect.ernet.in}
\affiliation{Department of Physics, Regional Engineering College,
Tiruchirapalli - 620 015, India}
\author{
M.~Lakshmanan%
}
\email{lakshman@bdu.ernet.in}
\affiliation{Centre for Nonlinear Dynamics, Department of Physics
Bharathidasan University, Tiruchirapalli 620 024, India}
\date{ }

\begin{abstract}

The conditions for synchronization in unidirectionally coupled chaotic
oscillators are revisited. We demonstrate with typical examples that
the conditional Lyapunov exponents (CLEs) play an important role in
distinguishing between intermittent and permanent synchronizations,
when the analytic conditions for chaos synchronization are not
uniformly obeyed. We show that intermittent synchronization can occur
when CLEs are very small positive or negative values close to zero
while permanent synchronization occurs when CLEs take sufficiently
large negative values. There is also strong evidence for the fact that
for permanent synchronization the time of synchronization is relatively
low while it is high for intermittent synchronization.

\end{abstract}
\maketitle

\section{Introduction}

Much attention has been focussed on the synchronization of chaotic
systems through different coupling schemes during the past decade or
so. It has potential applications not only in secure communication but
also in chaotic cryptography\cite{vaidya0}. In order to achieve
synchronization, coupled chaotic systems have to satisfy certain
conditions. According to Pecora and Carroll the coupled chaotic systems
can be regarded as master and slave systems which will perfectly
synchronize only if the sub-Lyapunov or conditional Lyapunov exponents
(CLEs) are all negative\cite{pecora1,pecora2}. However, recently there
are reports claiming that synchronization can be achieved even with
positive CLEs\cite{shuai,guemez,gutierrez}. Also there are situations
where one can observe intermittent or round-off induced synchronization
phenomenon\cite{baker,baker2,zhou} where locking occurs at certain time
intervals only, so that the synchronization of chaos need not be always
permanent. It is therefore important to understand clearly the
conditions under which chaos synchronization can occur and to know how
to distinguish between permanent and intermittent synchronizations
exhibited by coupled chaotic systems.

In this paper, the analytic conditions for synchronization in coupled
chaotic systems are revisited to show the difficulties in
distinguishing between intermittent and permanent synchronizations
which we illustrate by using simple chaotic systems. We demonstrate
that the conditional Lyapunov exponents (CLEs) play an important role
in distinguishing between intermittent and permanent synchronization,
when the analytic conditions for chaos synchronization are not
uniformly obeyed. We also point out that intermittent synchronization
can occur when the largest CLE has a value close to zero (either
positive or negative) while permanent synchronization occurs for
sufficiently large negative values of the CLEs.

In \Sec{sec2}, we describe briefly the notion of chaos synchronization
and the analytic conditions for synchronization which are to be
uniformly obeyed as well as the existence of negative conditional
Lyapunov exponents, with reference to a simple coupled system of
autonomous Duffing-van der Pol (ADVP) oscillators\cite{lakshman,
murali1}. \Sec{sec:cles} describes how the CLEs can be used to analyse
the nature of synchronization in coupled ADVP oscillators.  In
\Sec{sec3}, we show how conditional Lyapunov exponents play an
important role in distinguishing between the intermittent and permanent
synchronizations occurring in a class of unidirectionally coupled
nonlinear oscillators including coupled chaotic pendula\cite{baker},
coupled Duffing oscillators\cite{baker} and coupled
Murali-Lakshmanan-Chua (MLC) circuits\cite{lakshman} either in the
absence of the analytic criteria or when they are not uniformly obeyed.
We also show how the time of synchronization can be used qualitatively
to check the presence or absence of intermittent synchronization.
\Sec{sec4} summarizes the results.

\section{Chaos synchronization and the conditions}
\label{sec2}

In this section, we consider a set of two unidirectionally coupled
autonomous Duffing-van der Pol (ADVP) oscillators as a model system to
analyse the concept of chaos synchronization. Further, we show that the
derived analytic conditions for the occurrence of chaos synchronization
are uniformly obeyed in this system for a specific set of parametric
values (that is, the analytic conditions are shown to be valid for all
values of the variables, say, ($x_1, x_2, x_3) \in {\cal R}^3$ for all
$t$). We also indicate that when the usual analytic conditions are not
uniformly obeyed by a given system, then it may either exhibit
intermittent or permanent synchronization\cite{baker}. In this case,
the nature of the conditional Lyapunov exponents can be used profitably
to distinguish between intermittent and permanent synchronization.
\begin{figure}[!ht]
\includegraphics[width=\linewidth]{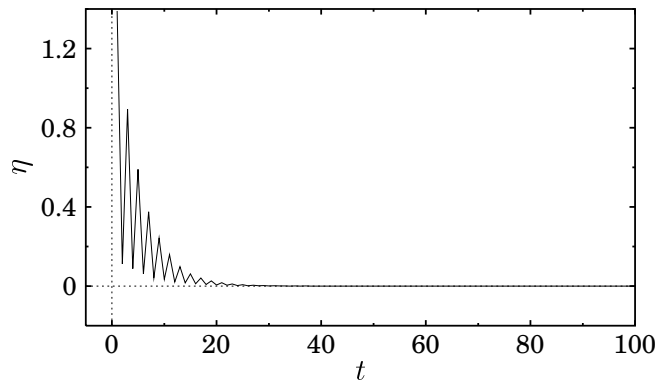}

\caption{Figure showing the variation of synchronization error $\eta$
as time elapses in ADVP oscillators [\Eqns{eqn:advp:1}] for $\nu=100$,
$\alpha=0.35$, $\beta=300$ and $\varepsilon = 1.35$.}

\label{ch5_f2}
\end{figure}

We consider the ADVP model in which Murali and Lakshmanan demonstrated
chaos synchronization \cite{lakshman,murali1}. The rescaled and
dimensionless version of the unidirectionally coupled ADVP oscillators
can be written as
\begin{subequations}
\label{eqn:advp:1}
\bear
\dot x_m & = & -\nu (x_m^3-\alpha x_m-y_m),\nonumber \\
\dot y_m & = & x_m-y_m-z_m, \nonumber \\
\dot z_m & = & \beta y_m ,
\label{eqn:advp:1a} \\
\nonumber \\
\dot x_s & = & -\nu (x_s^3-\alpha x_s-y_s) + \nu\varepsilon(x_m-x_s),\nonumber \\
\dot y_s & = & x_s-y_s-z_s, \nonumber \\
\dot z_s & = & \beta y_s.
\label{eqn:advp:1b}
\eear
\end{subequations}
Here $\nu$, $\alpha$, and $\beta$ are rescaled parameters, which are
fixed at $\nu=100$, $\alpha=0.35$ and $\beta=300$. One can define a
measure of the synchronization error, $\eta$, as
\begin{equation}
\eta = \sqrt{(x_m-x_s)^2+(y_m-y_s)^2+(z_m-z_s)^2}.
\label{eqn:eta}
\end{equation}
For synchronization, the above measure $\eta\to 0 $ as $t\to\infty$.
The synchronization error $\eta$ versus time is plotted in
\Fig{ch5_f2}.  The falling up of $\eta$ to zero is an indication of
synchronization at a finite time. However, this alone does not ensure
that the synchronization is permanent and one has to verify additional
criteria.  For this purpose, we will first discuss conditions for which
one can achieve permanent synchronization.

\renewcommand{\theenumi}{(\roman{enumi})}
\begin{enumerate}
\item

The criterion introduced by Fujisaka and Yamada \cite{fuji1, fuji2} for
high quality synchronization requires that the largest eigenvalue of
the Jacobian matrix corresponding to the flow evaluated on the
synchronization manifold be negative. In order to check this for the
system (\ref{eqn:advp:1}), let us consider the specific choice
$\varepsilon = 1+\alpha = 1.35$ in \Eqns{eqn:advp:1}. In this case, then one
can write the difference system of the ADVP oscillators for
$x^*=(x_m-x_s)$, $y^*=(y_m-y_s)$, $z^*=(z_m-z_s)$ in matrix form as
\begin{equation}
\left(
\begin{array}{c}
\dot x^* \cr \dot y^* \cr \dot z^*
\end{array}
\right)
=
\left(
\begin{array}{ccc}
-\nu (1+a) & \nu & 0 \cr
1 & -1 & -1 \cr
0 & \beta & 0
\end{array}
\right)
\left(
\begin{array}{c}
x^* \cr y^* \cr z^*
\end{array}
\right),
\label{eqn:advp:3}
\end{equation}
where $a = (x_m-x_s)^2+3x_mx_s \ge 0$. The
slave system (\ref{eqn:advp:1b}) synchronizes perfectly with the master
system (\ref{eqn:advp:1a}) only if all the eigenvalues of the above
linear system (\ref{eqn:advp:3}) possess negative real parts. One can
easily prove that this is indeed the case for (\ref{eqn:advp:3}).

\item

Recently, He and Vaidya~\cite{vaidya} developed a criterion for chaos
synchronization based on the notion of asymptotic stability of
dynamical systems, which refers to the condition for a given chaotic
system with master-slave configuration to reach the same eventual state
at a fixed (but sufficiently far enough) time irrespective of the
choice of initial conditions. One of the practical ways to establish
the asymptotic stability of the response subsystem is to find a
suitable Lyapunov function which can be defined as the square of the
magnitude of the vector describing the distance from the
synchronization manifold. Then the condition for all the perturbations
to decay to the synchronization manifold, without transient growth, is
that the time rate of the Lyapunov function has a negative magnitude
for all times\cite{gauthier}.

Now for the difference system (\ref{eqn:advp:3}), one can analytically
write a Lyapunov function in the following form for 
$\beta, \nu > 0$\cite{lakshman,murali1},
\begin{equation}
E = \frac{\beta}{2}\left[x^{*2}+\nu y^{*2}\right]+\frac{\nu}{2}z^{*2} \ge 0.
\end{equation}
Then the rate of change of $E$ along the trajectories is given by
\begin{equation}
\dot E = -\nu\beta \left [ ax^{*2} +(x^*-y^*)^2 \right] \le 0. 
\end{equation}
Since $E$ is a positive definite function and $\dot E$ is negative
definite for $(x^*, y^*, z^*)\in {\cal R}^3$, according to Lyapunov
theorem $x^*$, $y^*$ and $z^* \to 0$ as $t\to \infty$.  Thus perfect
synchronization occurs as $t \to \infty$.

\item

In addition, if all the sub-Lyapunov or conditional Lyapunov exponents
(CLE) are negative then one \emph{may} have perfect synchronization
between the master and slave systems. The CLEs are the corresponding
Lyapunov exponents of the slave system. For the coupled ADVP
oscillators [\Eqns{eqn:advp:1}] the CLEs can be calculated numerically
using the standard algorithm\cite{wolf}. We find that for the system
(\ref{eqn:advp:1}) with $\varepsilon = 1+\alpha = 1.35$, the numerical
value of the largest CLE is $-0.3134$. Thus, one may expect that the
ADVP oscillators will synchronize perfectly for $\varepsilon = 1.35$
which is indeed true.
\end{enumerate}
Consequently the conditions for synchronization in coupled chaotic
systems may be atleast any one of the following:
\begin{enumerate}

\item 
\label{syn:cond:1}

The largest eigenvalue of the Jacobian matrix corresponding to the flow
evaluated on the synchronization manifold \emph{must always be
negative} [see \Eqn{eqn:advp:3}].

\item 
\label{syn:cond:2}

Existence of a suitable Lyapunov function for the difference system
as discussed earlier.

\item
\label{syn:cond:3}

The sub-Lyapunov exponents or CLEs are all negative.

\end{enumerate} 
Among these three conditions, it is obvious that either of the
conditions \ref{syn:cond:1} and \ref{syn:cond:2} is both necessary and
sufficient for perfect or permanent synchronization as the very
definition of the latter implies these conditions. On the other hand
the condition \ref{syn:cond:3} is only a necessary one, because the
CLEs pertain to finite time averages only. As we have discussed above
for the specific parametric ($\varepsilon = 1.35$) case of the coupled
ADVP oscillators (\ref{eqn:advp:1}), all the conditions
\ref{syn:cond:1} - \ref{syn:cond:3} are satisfied and so pemanent
synchronization indeed occurs. However, suppose for a given system, if
the first two conditions are not uniformly obeyed and only the third
condition is satisfied, the question is whether the negativity of CLEs
alone ensures permanent synchronization.

In fact, recent studies show that in certain specific systems the
observed synchronization is \emph{not always} a perfect one and rather
one can have an \emph{intermittent synchronization}\cite{baker, muru1}
even when all the CLEs are negative. Further, there are reports stating
that one can achieve synchronization in the presence of positive
conditional Lyapunov exponents as well\cite{shuai}. So one has to
clarify the conditions under which permanent synchronization can occur
in coupled chaotic systems and how one can distinguish between the
intermittent and permanent synchronization.

\section{Nature of CLEs in ADVP oscillators
[E\lowercase{qs.}~(\ref{eqn:advp:1})] for general $\varepsilon$}
\label{sec:cles}
Based on the analysis of the previous section, one understands that the
coupled ADVP oscillators (\ref{eqn:advp:1}) show perfect
synchronization for $\varepsilon=1.35$. Now what happens to other
values of $\varepsilon$ in the coupled system (\ref{eqn:advp:1})? It
does not seem to be feasible either to deduce analytically the nature
of the largest eigenvalue of the Jacobian corresponding to the flow
evaluated on the synchronization manifold or to obtain explicitly a
suitable Lyapunov function.

\begin{figure}[!ht]
\begin{center}
\includegraphics[width=\linewidth]{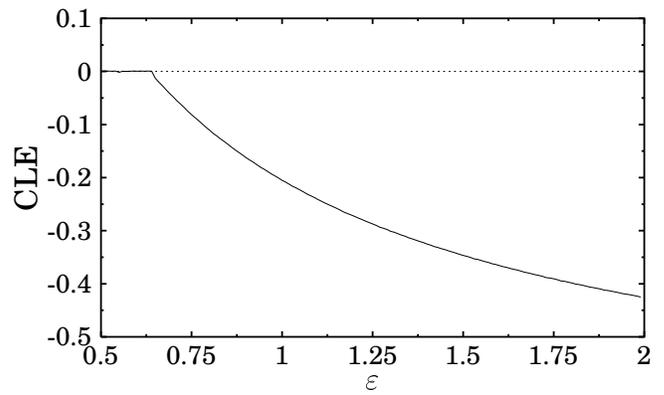}
\end{center}

\caption{Figure showing the variation of largest CLE as a function of
coupling strength ($\varepsilon$) for ADVP oscillators.}

\label{advp_lyex}
\end{figure}

In order to understand the nature of synchronization in
\Eqns{eqn:advp:1} for $\varepsilon\ne 1.35$, let us calcuate the CLEs.
\Fig{advp_lyex} shows the variation of the largest CLE as a function of
the coupling strength ($\varepsilon$). From the figure, one may expect
that the ADVP oscillators will synchronize for $\varepsilon \ge 0.65$.
However, the largest CLE value is almost zero in the range of
$\varepsilon\in (0.5, 0.65)$. On careful observation, by examining the
synchronization error $\eta$ with the addition of a small amount of
noise at every integration step, one finds that perfect synchronization
will occur in the coupled ADVP oscillators for $\varepsilon \ge 0.65$
only.

In addition, the Jacobian matrix corresponding to the flow evaluated on
the synchronization manifold can be given by
\begin{equation} 
\label{eq:advp:jacob}
J = \left( \begin{array}{ccc} -\nu
(3x_{\text{sm}}^2+c) & \nu & 0 \cr 1 & -1 & -1 \cr 0 & \beta & 0
\end{array} \right), 
\end{equation} 
where $\varepsilon = \alpha+c$ ($c=\text{constant}$), $\alpha = 0.35$. 
Then the eigenvalues take the form
\begin{eqnarray} 
\lambda_{1,2} & = &-\frac{1+\nu(3x_{\text{sm}}^2+c)}{2} \nonumber \\
 & &\pm\frac{\sqrt{[1+\nu(3x_{\text{sm}}^2+c)]^2
-4\nu(3x_{\text{sm}}^2+c-1)}}{2}, \nonumber\\
 \lambda_3 & = & 0.
\end{eqnarray} 
In the above, the sign of the largest eigenvaule is determined by the
factor $(3x_{\text{sm}}^2+c-1)$ inside the square root sign. When $c=1$
(that is, $\varepsilon = 1.35$), the largest eigenvalue is negative as
discussed earlier. Further, it is evident that the largest eigenvalue
remains negative for $c>1$.

On the other hand, if $c < 1$ then one has to evaluate the eigenvalues
numerically in order to check the nature of synchronization. For this
purpose we examine the variation of the eigenvalues numerically.
\begin{figure}[!ht] 
\begin{center}
\includegraphics[width=\linewidth]{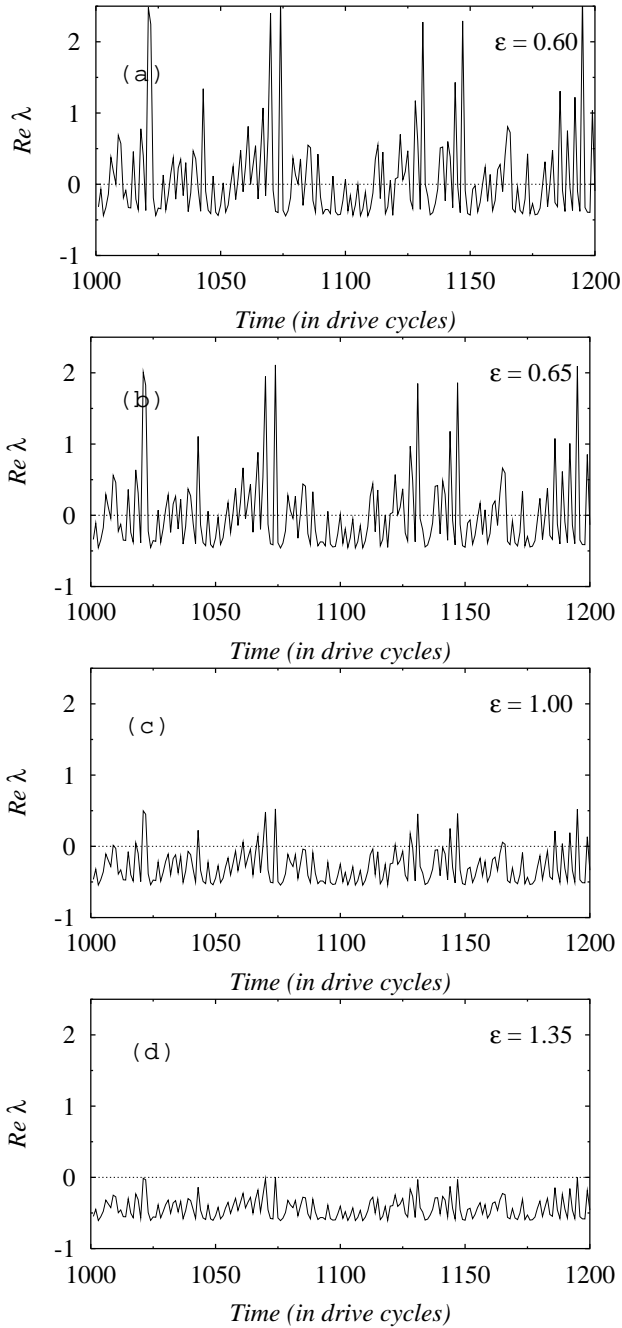}
\end{center}
\caption{Figure showing the plot of the largest eigenvalue of the
Jacobian matrix (\ref{eq:advp:jacob}) versus time for ADVP oscillators:
(a) $\varepsilon=0.60$, (b) $\varepsilon=0.65$, (c) $\varepsilon=1.00$
and (d) $\varepsilon=1.35$.}

\label{advp_lf}
\end{figure}
\Figs{advp_lf} show the variation of the largest eigenvalue as a
function of time. \Fig{advp_lf}(a) illustrates the variation of the
largest eigenvalue for $\varepsilon=0.60$ ($c=0.25$). Here the largest
eigenvalue oscillates chaotically about zero with an average of the
order of $+10^{-1}$ and hence no synchronization is possible for
$\varepsilon=0.60$. For $\varepsilon=0.65$, the largest eigenvalue
still oscillates chaotically about zero [see \Fig{advp_lf}(b)] with an
average $\approx 10^{-2}$. In the latter case the largest eigenvalue
spends more time in the negative region despite the fact that its
average value is positive. In addition, for $\varepsilon=0.65$, the CLE
has a value of $-0.01298$ and thus it shows permanent synchronization.
Similar arguments hold good for $0.65 < \varepsilon < 1.35$.
\Figs{advp_lf}(c) and \ref{advp_lf}(d) depict the largest eigenvalue
for $\varepsilon=1.00$ and $\varepsilon=1.35$, respectively.

In addition to the above, one can also verify the condition
\ref{syn:cond:2}, by evaluating the time rate of the Lyapunov function
\begin{figure}[!ht] 
\begin{center}
\includegraphics[width=\linewidth]{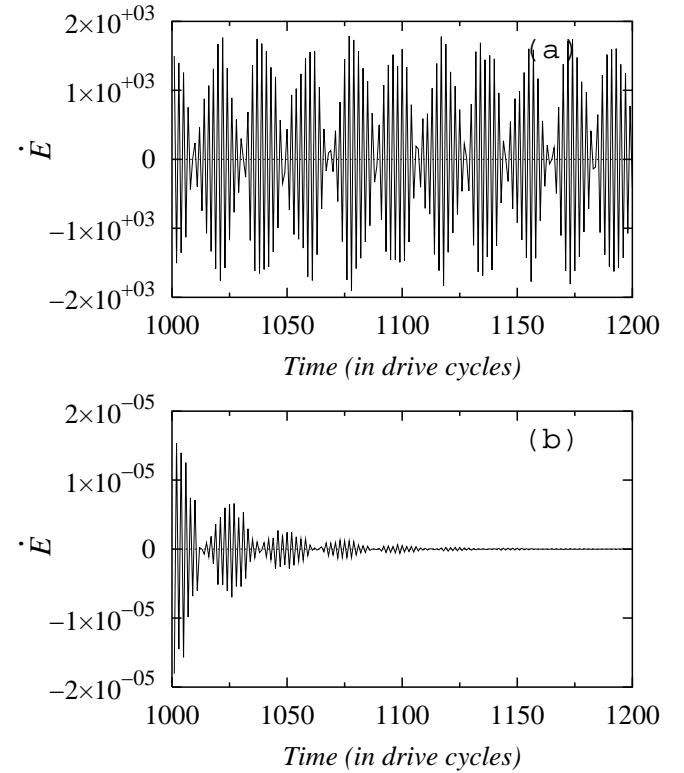}
\end{center}
\caption{Figure showing the plot of $\dot E$ in the ADVP oscillators
[Eq.~\ref{eq:advp:edot}] as a function of time for (a)
$\varepsilon=0.64$ and (b) $\varepsilon=0.65$.}
\label{advp_lyap}
\end{figure}
corresponding to the square of the magnitude of the vector describing
the distance from the synchronization manifold\cite{gauthier}, which is
given by
\begin{eqnarray}
\label{eq:advp:edot}
\dot E & = & -\nu(3x_{\text{sm}}^2+c)x^{*2}-y^{*2} \nonumber \\
& & +(1+\nu)x^*y^*+(\beta+1)y^*z^* \le 0. 
\end{eqnarray}
\Fig{advp_lyap}(a) shows the variation of $\dot E$ for
$\varepsilon=0.64$. From the figure it is clear that the condition
\ref{syn:cond:2} is not uniformly satisfied. Thus, there is no
synchronization in the coupled ADVP oscillators (\ref{eqn:advp:1}) for
$\varepsilon=0.64$. However, $\dot E\to 0$ at a finite time for
$\varepsilon=0.65$ and $\varepsilon>0.65$ [see \Fig{advp_lyap}(b)] and
hence perfect synchronization does occur in the coupled ADVP
oscillators for $\varepsilon\ge 0.65$.

Thus from the above analysis and from the nature of the CLEs, one may
conclude that perfect synchronization in the coupled ADVP oscillators
(\ref{eqn:advp:1}) will occur for $\varepsilon \ge 0.65$. The role of
CLEs become even more important when both the analytic conditions
\ref{syn:cond:1} and \ref{syn:cond:2} are not uniformly satisfied. In
such cases, in the following, we will analyse how the nature of CLEs
can be effectively used to understand synchronization in typical
coupled chaotic systems.

\section{Distinguishing intermittent and permanent synchronization
using CLEs}

\label{sec3}

We now wish to clarify the conditions for which one can observe
intermittent and permanent synchronization, albeit in a qualitative
way, when the analytic conditions \ref{syn:cond:1} and \ref{syn:cond:2}
are not uniformly obeyed. In this regard we wish to analyse the nature
of synchronization on the basis of the conditional Lyapunov exponents
(CLEs) by considering three typical dynamical systems as examples. They
are namely (i) coupled chaotic pendula, (ii) coupled Duffing
oscillators and (iii) coupled MLC circuits. We find that all these
three systems do not satisfy the analytic conditions \ref{syn:cond:1}
and \ref{syn:cond:2} uniformly in contrast to the coupled ADVP
oscillators considered in Sec.~\ref{sec2}. We demonstrate how the
nature of the CLEs become important in distinguishing the intermittent
and permanent synchronizations exhibited by these systems.

\subsection{Coupled chaotic pendula}
\begin{figure*}[!ht]
\includegraphics[width=0.8\linewidth]{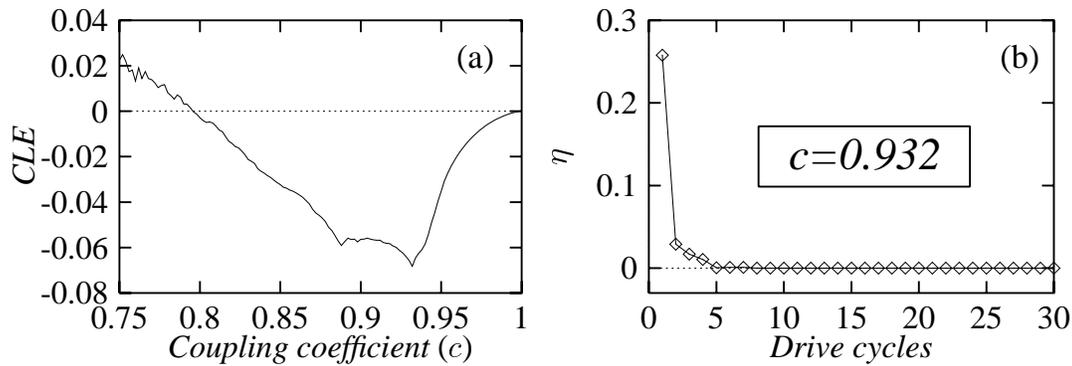}

\caption{(a) Variation of the conditional Lyapunov exponent (CLE) as a
function of $c$ and (b) synchronization error $\eta$ [see
\Eqn{eqn:pen:eta}] versus time in drive cycles for coupled pendula.}

\label{fig1}
\end{figure*}

First we consider a pair of coupled chaotic pendula defined by the
following set of equations and investigated by Baker, Blackburn and
Smith\cite{baker}
\begin{subequations}
\label{eq:pen1}
\bear
&&\ddot\theta_m+Q^{-1}\dot\theta_m =  \Gamma_0 \cos \Omega t ,
\label{eq:pen1a}\\
&&\ddot\theta_s+Q^{-1}\dot\theta_s  =  \Gamma_0 \cos \Omega t 
+ c[\sin\theta_s - \sin\theta_m].
\label{eq:pen1b}
\eear
\end{subequations}
Here $\theta_m$ and $\theta_s$ correspond to the angular coordinates of
the master and slave systems, respectively. $Q^{-1}$ corresponds to the
damping factor, $\Gamma_0$ is the normalized drive torque, $\Omega$ is
the drive frequency and $c$ is the coupling strength. By fixing the
parameters at $Q=5.0$, $\Gamma_0=1.2$ and $\Omega=0.5$, the quality of
synchronization in the above coupled system (\ref{eq:pen1}) has been
studied. This has been facilitated by finding the synchronization error,
\begin{equation}
\eta = \sqrt{(\theta_s-\theta_m)^2+(\dot\theta_s-\dot\theta_m)^2}.
\label{eqn:pen:eta}
\end{equation}
As seen earlier, for synchronization $\eta\to 0$ as $t\to\infty$.

It has been shown by Baker, Blackburn and Smith\cite{baker} that
permanent synchronization in the above coupled identical chaotic
pendula (\ref{eq:pen1}) does not occur except as a numerical artifact
arising from finite computational precision. Further, they showed that
the synchronization of the above coupled pendula (\ref{eq:pen1}) is
always {\em intermittent}, for any value of the coupling coefficient,
by using numerical and analytical tests. However, contrary to the above
the present authors have reported that there exists atleast certain
range of coupling coefficient values for which one can observe
permanent synchronization by computing the CLEs\cite{muru1,muru2}. In
order to understand this we proceed as follows.
\begin{figure*}[!ht]
\includegraphics[width=\linewidth]{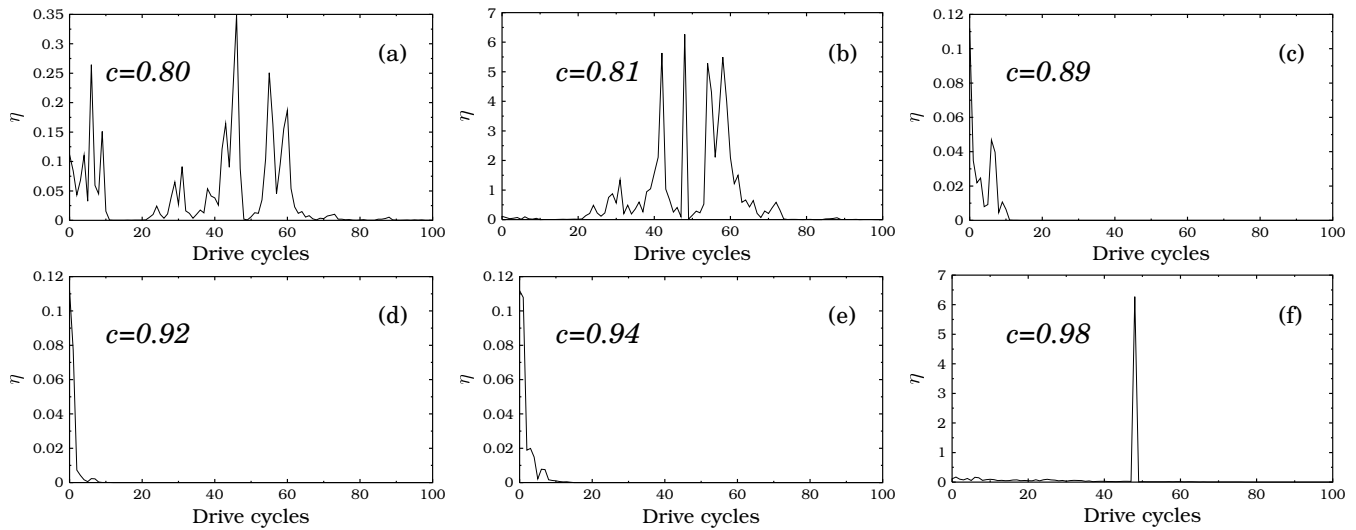}

\caption{Figure showing the variation of synchronization error $\eta$
versus time  in the coupled chaotic pendula for (a) $c=0.80$, (b)
$c=0.81$, (c) $c=0.89$, (d) $c=0.92$, (e) $c=0.94$ and (f) $c=0.98$.
Fluctuation in $\eta$ shows the possibility of having intermittent
synchronization. In calculating $\eta$, tiny noise levels ($\approx
10^{-15}$) are included at each time step to avoid round off induced
synchronization.}

\label{pen_eta}
\end{figure*}

For the coupled chaotic pendula (\ref{eq:pen1}), as shown in
Ref.~\cite{baker} the eigenvalues of the Jacobian matrix corresponding
to the flow evaluated on the synchronization manifold take the form
\begin{eqnarray}
\lambda_{1,2} & = & \frac{-1}{2Q}\left[
1\pm\sqrt{1-2Q^2(1-c)\cos\theta_{\text{sm}}}
\right], \nonumber\\
\lambda_3 & = & 0.
\end{eqnarray}
Here $\theta_{\text{sm}}$ corresponds to the angle coordinate on the
synchronization manifold. On careful numerical observation, it has been
shown by Baker, Blackburn and Smith~\cite{baker} that the term inside
the square root sign varies chaotically about unity with an average
value which is slightly less than one for a range of $c$ values. This
implies that the largest eigenvalue is not always less than zero and
hence the condition \ref{syn:cond:1} is not obeyed uniformly.

Further, calculating the Lyapunov function which is the square of the
magnitude of the vector describing the distance from the
synchronization manifold\cite{gauthier}, the sufficient condition for
permanent synchronization can be written as\cite{baker}
\begin{equation}
\dot E = \left (1-c-\cos\theta_{\text{sm}}\right )\theta^{*}\omega^{*}
-Q^{-1}\omega^{*2} \le 0,
\end{equation}
where $\omega = \dot\theta$, $\theta^{*} = \theta_{m}-\theta_{s}$ and
$\omega^{*} = \omega_{m}-\omega_{s}$. The time series of the above
expression varies intermittently about zero with an average slightly
less than zero for a range of $c$ values [$c\in (0.79,1.0)$] and hence
the condition \ref{syn:cond:2} is also not obeyed for the coupled
chaotic pendula.  Thus, in the light of the above evidence, Baker,
Blackburn and Smith~\cite{baker,baker2} conclude that intermittent
synchronization can be a plausible behaviour in the coupled pendulum as
the locking occurs intermittently. In order to verify this assertion,
we have carried out an analysis\cite{muru1} by calculating the
conditional Lyapunov exponents of system (\ref{eq:pen1}) in the
parameter range $c\in (0.75,1.0)$. We find that there exists a range of
$c$ values for which one can indeed have permanent synchronization,
where Baker, Blackburn and Smith\cite{baker} have expected intermittent
synchronization. In our calculations we have used the standard Wolf et
al algorithm\cite{wolf} and used $5000$ drive cycles for calculations
after leaving out $5000$ drive cycles as transient.

Figure \ref{fig1}(a) shows the variation of CLE as a function of $c$
for the coupled pendula and it takes negative values for $0.796 \le c
<1.0$ only. The CLE value for $c=0.79$ is $+0.00320$ ($\approx
10^{-3}$), for which synchronization can not occur and hence the
observed intermittent synchronization (see Fig.~1 in Ref~\cite{baker})
is a computer artifact. However, we find that for $c=0.932$, the CLE
value becomes the lowest ($-0.06825$) and it is relatively a large
negative value for which intermittent synchronization is absent as
shown in \Fig{fig1}(b). The absence of the intermittent synchronization
has been verified upto $2\times 10^6$ drive cycles even with the
addition of tiny noise levels showing that permanent synchronization
does occur for this value of $c$. The same phenomenon persists over a
range of $c$ values close to $0.932$. \Figs{pen_eta} show the variation
of the synchronization error for various $c$ values. From the figures
it is clear that one can indeed have intermittent synchronization for a
range of $c$ values as noted by Baker, Blackburn and
Smith\cite{baker}.  For example, the system definitely exhibits
intermittent synchronization for $c=0.80$, $0.81$ and $0.98$, where we
find that in this range of $c$ values (that is, $c\le 0.88$ or $c\ge
0.95$) the largest CLEs take values very close to zero (either positive
or negative). But more interestingly we also find that there exists
another range of $c$ values from $0.89$ to $0.94$ where one can have
permanent synchronization. In calculating $\eta$ for
\Figs{pen_eta}(a)-(f), we have added a small amount of noise ($\approx
10^{-15}$) at each time step to avoid the effect of round off induced
\begin{figure}[!h]
\includegraphics[width=\linewidth]{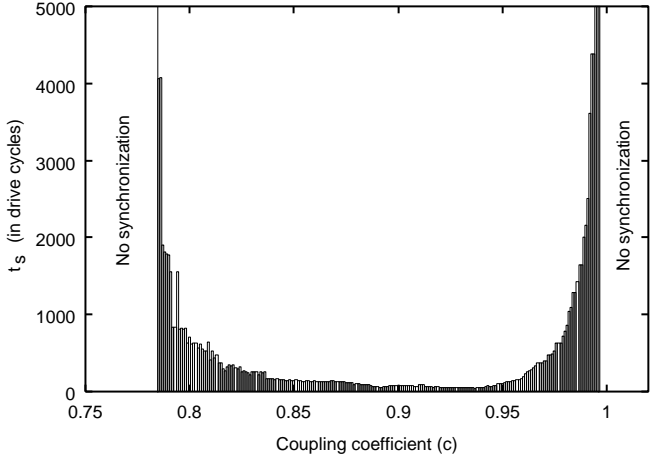}
\caption{Figure showing the variation of time of synchronization 
$t_s$ (in drive cycles) versus coupling strength for coupled chaotic
pendula.}
\label{pen_ts}
\end{figure}
synchronization. We note that the CLEs are negative and the largest CLE
is relatively away from zero in this range. Thus we note that the
actual values of the CLEs, including the largest CLE, seem to
distinguish between permanent and intermittent synchronization.

Next we calculate the time of synchronization ($t_s$), that is, the
time taken by the system for which the synchronization error becomes
zero. \Fig{pen_ts} depicts the variation of $t_s$ as a function of the
coupling coefficient. It can be easily seen that the value of $t_s$ is
very low for the range of $c$ values where the CLEs are large negative
which corresponds to permanent synchronization. Similarly, the value of
$t_s$ becomes high in the regions where the largest CLE has a value
close to zero (either negative or positive) which corresponds to
intermittent synchronization, as there exists large fluctuations in the
finite time Lyapunov exponents here.

It is clear from the above analysis on the coupled chaotic pendula
(\ref{eq:pen1}) that there exists intermittent synchronization for
certain ranges of coupling strength values. However, permanent
synchronization does occur for certain other range of $c$ values. In
particular, we have pointed out that when the CLEs become very close to
zero (either positive or negative) one can have intermittent
synchronization while permanent synchronization does occur when the
CLEs become large negative. Thus the CLEs play a crucial role in
distinguishing between intermittent and permanent synchronization. In
the following subsections we will examine the existence of similar
intermittent synchronization in the coupled Duffing equations and
coupled MLC circuits.

\subsection{Coupled Duffing oscillators}

Now we consider the case of two coupled Duffing oscillators described
by the following set of equations,
\begin{subequations}
\bear
&&\ddot x_m + \alpha \dot x_m + \beta x_m^3 = f\cos \omega t, \\
&&\ddot x_s + \alpha \dot x_s + \beta x_s^3 = f\cos \omega t + c(x_s - x_m),
\eear
\end{subequations}
where $\alpha = 0.1$, $\beta = 1.0$, $f=10$ and $\omega=1.0$ and $c$ is
the coupling strength. By setting $\dot x = y$, $z = \omega t$, one can 
rewrite the above equations as a set of first order ordinary differential
equations of the form
\begin{subequations}
\label{duf:syn1} 
\bear
\dot x_m & = & y_m \nonumber\\
\dot y_m & = & -\alpha y_m -\beta x_m^3 + f\cos z_m \nonumber\\
\dot z_m & = & \omega
\\ \nonumber \\
\dot x_s & = & y_s \nonumber\\
\dot y_s & = & -\alpha y_s -\beta x_s^3 + f\cos z_s + c(x_s - x_m) \nonumber\\
\dot z_s & = & \omega
\eear
\end{subequations}

First, we study the nature of synchronization in the above coupled
Duffing equations by analysing the conditions \ref{syn:cond:1} and
\ref{syn:cond:2} of Sec.~\ref{sec2} as in the case of chaotic pendula
\begin{figure}[!ht]
\includegraphics[width=0.9\linewidth]{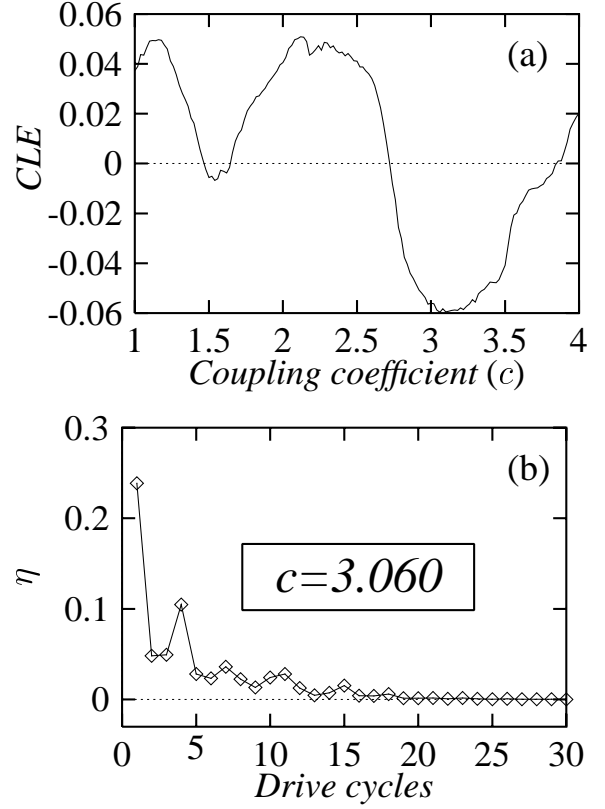}
\caption{
(a) Variation of the conditional Lyapunov exponent (CLE) as a function
of $c$ and (b) synchronization error $\eta$ for coupled Duffing oscillators.
}
\label{duf_cfig}
\end{figure}
discussed above.  Here the eigenvalues of the Jacobian matrix
corresponding to the flow evaluated on the synchronization manifold can
be explicitly written as
\begin{equation}
\lambda_{1,2} = \frac{-\alpha\pm\sqrt{\alpha^2 
+ 4(3\beta x_{\text{sm}}^2+c)}}{2\alpha},\;\;\;\;
\lambda_3 = 0,
\end{equation}
where $x_{\text{sm}}$ denotes the $x$-variable evaluated on the
synchonization manifold. We found that the largest of eigenvalue
oscillates chaotically about zero with an average value very close to
\begin{figure*}[!ht]
\includegraphics[width=0.9\linewidth]{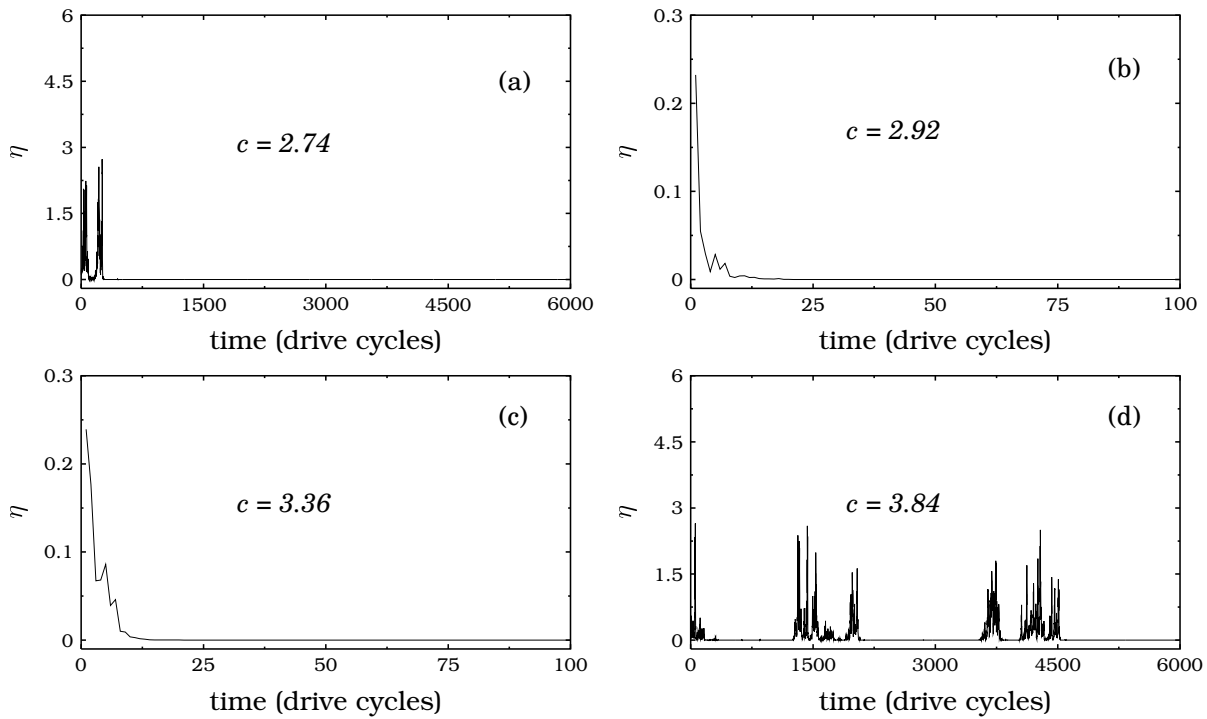}
\caption{Figure showing the variation of synchronization error $\eta$
versus time  in the coupled Duffing oscillators for (a) $c=2.74$, (b)
$c=2.92$, (c) $c=3.36$ and (d) $c=3.84$ with the inclusion of tiny noise 
levels of the order of $10^{-15}$ at each integration step.}
\label{duf_eta}
\end{figure*}
zero ($\approx\pm 10^{-3}$) for two ranges of $c\in (1.48,1.64)$ and $c\in
(2.74,3.84)$. Thus the condition \ref{syn:cond:1} is not uniformly
obeyed. In addition, one can write the condition \ref{syn:cond:2} as
\begin{equation}
\dot E = (1-3\beta x_{\text{sm}}^2-c)x^{*}y^{*} -\alpha y^{*2} \le 0,
\end{equation}
with $x^{*} = x_m - x_s$ and $y^{*} = y_m - y_s$. On actual numerical
simulation, it has been found that the time series of the above
expression oscillates chaotically about zero with an average of the
order of $\pm 10^{-3}$. This confirms that the condition
\ref{syn:cond:2} is also not satisfied uniformly. Thus, the only way to
\begin{figure}[!h]
\includegraphics[width=\linewidth]{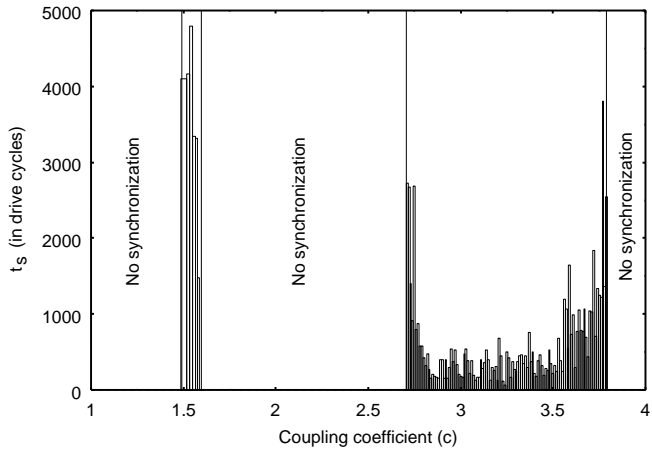}

\caption{Figures showing the variation of time of synchronization $t_s$
versus $c$ in the coupled Duffing oscillators.}

\label{ts_duf}
\end{figure}
understand the nature of synchronization in \Eqns{duf:syn1} is to
analyse the CLEs. We have again calculated the largest CLE as a
function of the coupling strength. \Fig{duf_cfig} shows a plot of the
largest CLE versus coupling strength ($c$). We find that the CLE takes
small negative values close to zero ($\approx 10^{-3}$) for
$c\in(1.48,1.64)$, where synchronization is essentially intermittent
and numerical artifact does arise in this range. In addition, there
exists another range of $c$ values, $2.74\le c\le 3.84$, where the CLEs
are again negative (cf.  \Fig{duf_cfig}(a)).  We find that for
$c=3.06$, the CLE takes the lowest value of $-0.05967$ (large negative
value) for which persistent synchronization occurs and the intermittent
synchronization is absent (cf. \Fig{duf_cfig}(b)).  The variation of
synchronization error for a range of selected $c$ values is depicted in
\Figs{duf_eta}(a)-(d). It can be noted from the figures that for
$c=2.74$ and $3.84$ (\Fig{duf_eta}(a) and (d)) the system can exhibit
intermittent synchronization while for $c=2.92$ and $3.36$
(\Fig{duf_eta}(b) and (c)) the system exhibits permanent
synchronization. Here also tiny noise levels in all the variables were
included at each integration step to ensure the absence of round off
induced synchronization.

As in the case of coupled pendula, we have calculated the time of
synchronization ($t_s$) by varying the coupling strength. \Fig{ts_duf}
shows the variation of $t_s$ as a function of the coupling strength for
this case. From the figure, it can be noted again that $t_s$ takes
relatively low values for a region of $c$ values where the CLEs have
large negative values which corresponds to permanent synchronization.
Similarly, the values of $t_s$ becomes high in the regions where the
largest CLE has a value close to zero or positive which corresponds to
intermittent synchronization.
\begin{figure*}[!ht]
\includegraphics[width=0.8\linewidth]{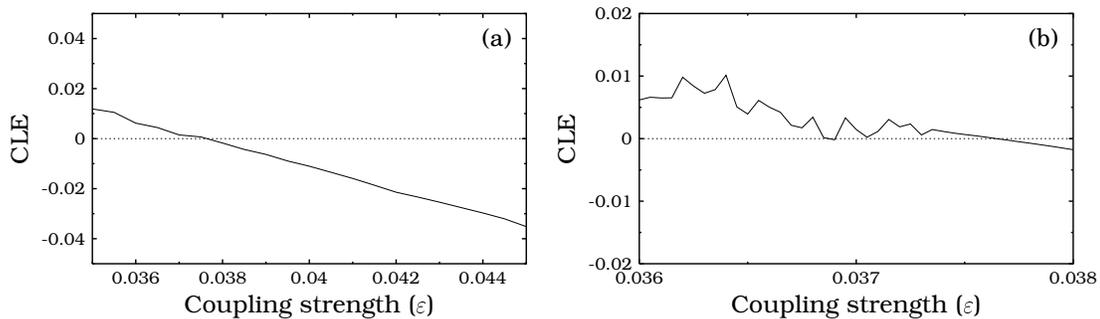}

\caption{Figure showing (a) the variation of largest CLE as a function
of the coupling coefficient $\varepsilon$ for the coupled MLC circuits
and (b) blow up of a small region in (a) indicating the fluctuations of
the largest CLE near zero.}

\label{mlc_lyex}
\end{figure*}
\begin{figure*}[!ht]
\includegraphics[width=\linewidth]{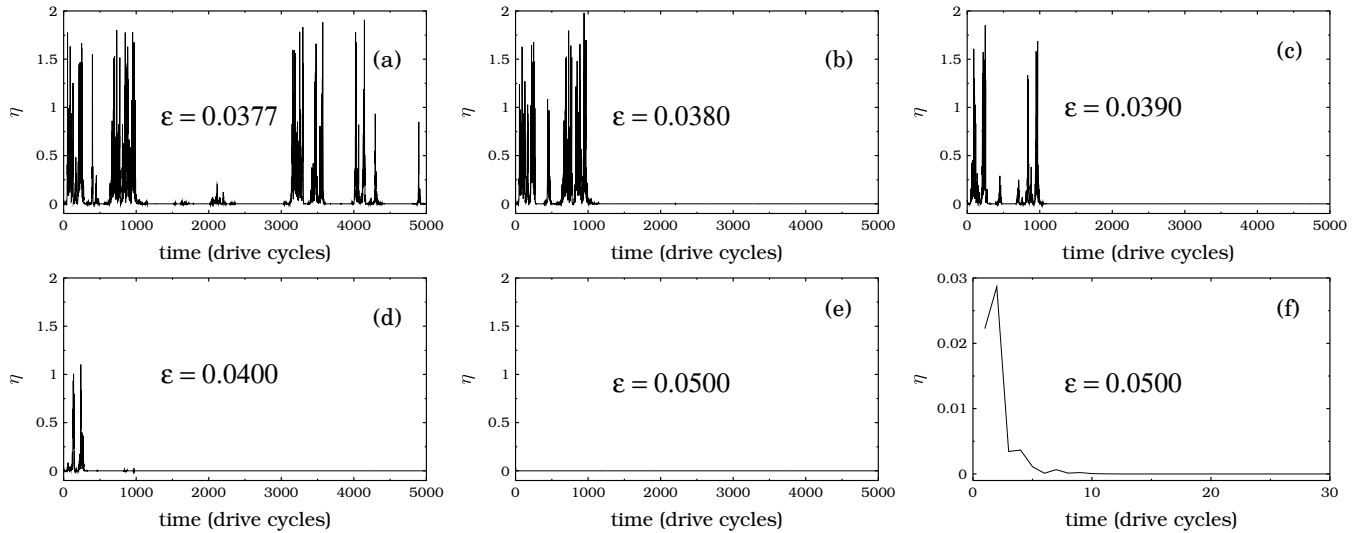}

\caption{Figure showing the variation of synchronization error $\eta$
versus time in the coupled MLC circuits for (a) $\varepsilon = 0.0377$,
(b) $\varepsilon = 0.0380$, (c) $\varepsilon = 0.0390$, (d)
$\varepsilon = 0.0400$, (e) $\varepsilon=0.0500$ and (f) a small region
in (e) with the inclusion of tiny noise levels at each integration
step.}

\label{mlc_eta}
\end{figure*}

\subsection{Coupled MLC circuits}

Finally we consider the case of the coupled Murali-Lakshmanan-Chua
(MLC) circuits, represented by the following set of
equations\cite{lakshman}
\begin{subequations}
\label{eqn:mlc}
\bear
\dot x_m & = & y_m-h(x_m), \nonumber \\
\dot y_m & = & -\beta x_m +\sigma y_m + F\sin z_m \nonumber \\
\dot z_m & = & \omega 
\eear
\bear
\label{eqn:mlc1}
\dot x_s & = & y_s-h(x_s)+\varepsilon (x_m-x_s), \nonumber \\
\dot y_s & = & -\beta x_s +\sigma y_s + F\sin z_s \nonumber \\
\dot z_s & = & \omega
\label{eqn:mlc2}
\eear
\end{subequations}
where $\varepsilon$ is the coupling strength and $h(x) = bx - 0.5(a-b)
(|x+1| - |x-1|)$. The uncoupled system of the above MLC circuits has
been shown to exhibit chaos for the choice of parameter values
$\beta=1$, $\sigma=1.015$, $a=-1.02$ and $b=-0.55$. We are interested
to analyse the existence of intermittent synchronization in this case
also.

First let us check the validity of the conditions \ref{syn:cond:1} and 
\ref{syn:cond:2}. In this case, the eigenvalues of the Jacobian matrix
of the flow evaluated on the synchronization manifold can be written as
\begin{eqnarray}
\lambda_{1,2} & = & -\frac{1}{2}
\left[h'(x_{\text{sm}})+\beta-\varepsilon\right]\nonumber \\
& & \pm\frac{1}{2}\sqrt{\left[h'(x_{\text{sm}})+\beta-\varepsilon\right]^2 
- 4[(h'-\varepsilon)\beta+\sigma]},\nonumber\\
\lambda_3 & = & 0.
\end{eqnarray}
Here $h'(x_{\text{sm}})$ corresponds to the derivative of $h(x)$ with
respect to $x$ variable evaluated on the synchronization manifold. From
numerical analysis, we have found that the largest of the above
eigenvalues oscillates chaotically about zero with an average slightly
less than zero for a range of $\varepsilon\in (0.036, 0.04)$ and thus
the condition \ref{syn:cond:1} is not uniformly obeyed. On the other
hand, the condition \ref{syn:cond:2} for the coupled MLC circuits can
be written as
\begin{equation}
\dot E = \left [\varepsilon-h'(x_{\text{sm}})\right ]x^{*2}
-\beta y^{*2}+(1-\sigma)x^{*}y^{*} \le  0,
\end{equation}
with $x^{*} = x_m - x_s$ and $y^{*} = y_m - y_s$. Again by numerical
analysis, we find that the above inequality is not uniformly obeyed and
hence the condition \ref{syn:cond:2} is not satisfied. However, the
coupled system (\ref{eqn:mlc}) has been shown to exhibit perfect
synchronization\cite{lakshman}. In order to analyse this we have
evaluated the CLEs as in the case of chaotic pendula and Duffing
oscillators.

The variation of the CLEs as a function of the coupling strength
($\varepsilon$) is shown in \Fig{mlc_lyex}. The largest CLE changes its
sign from positive to negative values at $\varepsilon = 0.0377$ and it
is slightly less than zero over the range $0.0377 < \varepsilon \lesssim
0.04$. In this range the synchronization completely depends on the
choice of initial conditions and the computer precision used. This
implies that the observed synchronization is essentially due to the
finite computational precision and accordingly locking ($\eta=0$)
occurs after relatively long time, leading to an intermittent
synchronization in this range of coupling strength.

We have also calculated the synchronization error ($\eta$) for various
values of coupling strength as a function of time as in the case of
previous systems and the results are depicted in \Fig{mlc_eta}. From
\Figs{mlc_eta}(a)-(d), it is evident that one can have intermittent
synchronization for $\varepsilon=0.0377$, $0.0380$, $0.0390$ and
$0.04$. \Fig{mlc_eta}(e) and \Fig{mlc_eta}(f) indicates the presence of
permanent synchronization for $\varepsilon=0.05$ where the sharp and
quick fall off in the synchronization error towards zero occurs (see
\Fig{mlc_eta}(f)).

\begin{figure}[!ht]
\includegraphics[width=\linewidth]{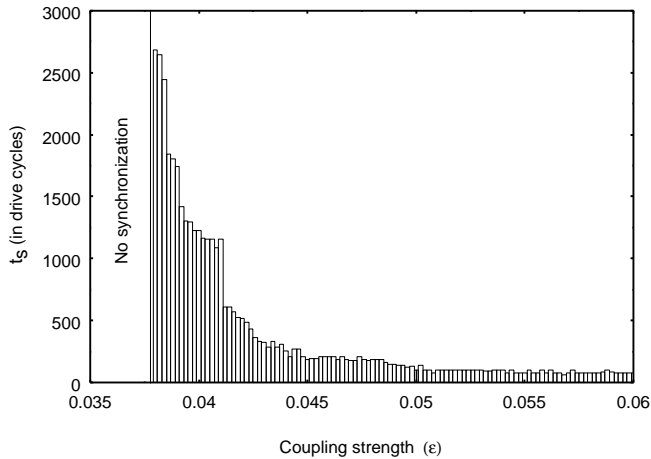}
\caption{Figure showing the variation of time of synchronization 
$t_s$ versus coupling strength $\varepsilon$ for coupled MLC 
circuits.}
\label{mlc_ts}
\end{figure}

\Fig{mlc_ts} shows the variation of time of synchronization ($t_s$) as
a function of the coupling strength ($\varepsilon$). As in the case of
coupled pendula and Duffing oscillators, $t_s$ takes relatively low
values for the range of $\varepsilon$ values where the CLEs take large
negative values corresponding to permanent synchronization. However,
the values of $t_s$ becomes high in the regions where the largest CLE
has a value close to zero or positive which corresponds to intermittent
synchronization. Thus one can conclude that permanent synchronization
in coupled MLC circuits (\ref{eqn:mlc}) occur for $\varepsilon >
0.04$.

\section{Summary and conclusions}
\label{sec4}

By considering typical examples of coupled oscillator models, we have
discussed the quality of synchronization using various analytical and
numerical tests. It has been noted that one can have an
\emph{intermittent} as well as \emph{permanent} synchronization in the
coupled oscillator systems depending on the choice of coupling
strength. We have pointed out that the Conditional Lyapunov Exponents
(CLEs) can be used to distinguish between the intermittent and
permanent synchronization when the other criteria for asymptotic
stability are not uniformly obeyed. Particularly, we find that
intermittent synchronization can occur when CLEs are very small
positive or negative values close to zero while persistent
synchronization occurs when CLEs take sufficiently large negative
values. This fact is further supported by the relative time taken for
the system to approach the synchronization manifold.

The present work is mainly concerned with the qualitative analysis of
chaos synchronization in unidirectionally coupled systems based on
conditional Lyapunov exponents. In order to understand the entire
dynamics one has to analyse the nature of the attractors that exist in
the coupled system and their bifurcations. It is obvious that we have
not made any quantitative analysis on \emph{how small} or \emph{large}
be the magnitude of the largest conditional Lyapunov exponent in
distinguishing the intermittent and permanent synchronizations. Work is
in progress along these directions.

\begin{acknowledgments}

This work was supported by Department of Science and Technology, 
Govt. of India and National Board for Higher Mathematics
(Department of Atomic Energy) in the form of research projects.

\end{acknowledgments}

\bibliography{pap2e}

\end{document}